\documentclass[12pt]{article}
\begin {document}

\begin {center}
{\bf {\Large Standing magnetic wave on Ising ferromagnet: Nonequilibrium phase transition.}}
\end {center}
\begin {center}{\bf{ $\bf Ajay Halder^\star$ and $\bf Muktish Acharyya^\dagger$}
\\{Department of Physics,}\\{Presidency University}\\{86/1 College Street, Kolkata-73, India}
\\{$\bf^\star ajay.rs@presiuniv.ac.in$}
\\{$\bf^\dagger muktish.physics@presiuniv.ac.in$}}
\end {center}
\vskip 0.2cm
\begin {abstract}
The dynamical response of an Ising ferromagnet to a plane polarised standing magnetic field 
 wave is modelled and studied here by {\it Monte Carlo simulation} in two dimensions.  
The amplitude of standing magnetic wave is modulated along the direction $x$.
 We have detected two main dynamical phases namely, {\it pinned} and {\it oscillating spin clusters}.
 Depending on the value of field amplitude the system is found to undergo a phase transition
 from oscillating spin cluster to pinned as the system is cooled down. The time averaged magnetisation over a 
full cycle of magnetic field oscillations is defined as the {\it dynamic order parameter}. 
The transition is detected by studying the temperature dependences 
of the variance of the dynamic order parameter, the derivative of the dynamic order parameter 
and the dynamic specific heat. The dependence of the transition temperature on the 
magnetic field amplitude and on the wavelength of the magnetic field wave is studied 
at a single frequency. 
A comprehensive phase boundary is drawn in the plane described by the temperature and field amplitude
 for two different wavelengths of the magnetic wave.
The variation of {\it instantaneous line magnetisation} % and {\it structure factor}
 during a period of magnetic field oscillation for standing wave mode
 is compared to those for the propagating wave mode. Also the probability that a spin at any site,
 flips, is calculated. The above mentioned variations and the probability of spin flip
clearly distinguish between the dynamical phases formed by propagating magnetic wave and by 
standing magnetic wave in an Ising ferromagnet.
\end {abstract}

\vskip 0.6cm


\newpage

{\noindent \bf I. Introduction}
\vskip 0.2cm

The study of the dynamical response of a {\it thermodynamical system} has become an active field
 of research \cite{rmp,ijmpc} in recent years. Ferromagnetic system is one of some  
important systems whose response to various kinds of driving force in equilibrium as well as in 
non-equilibrium situations hold the key attention of many researchers for a long time.
 A ferromagnetic system responses in a unique way to a {\it time dependent} magnetic field 
and studies of such {\it dynamical responses} revealed many interesting facts of some 
dynamical behaviour of the system. The nonequilibrium dynamic phase 
transition and the hysteretic response are the main characteristic features of the 
ferromagnetic system driven by time dependent magnetic field.
 Some observations or studies regarding -- 
{\it (i)} divergences of dynamic specific heat and relaxation time near
 transition point \cite{divCprelax,maphysica}, {\it (ii)} divergence of the relevant length scale 
near transition point \cite{divlnth}, {\it (iii)} studies regarding existence of
 tricritical point \cite{rev99,rev02}, {\it (iv)} its relation with stochastic resonance \cite{rev99},
and the hysteresis loss \cite{rev98} etc., establish that the dynamic phase transition
 is similar in many aspects to the well known equilibrium thermodynamic phase transition.
 This fact is further supported by some experimental findings like-- {\it (i)} detection 
of dynamic phase transition in the ultra-thin $Co$ film on $Cu (001)$ system
 by surface {\it magneto-optic Kerr effect} \cite{kerr1,kerr2}, {\it (ii)} direct excitation 
of propagating spin waves by focussed ultra short optical pulse \cite{excitspin}, 
{\it (iii)} the transient behaviour of dynamically ordered phase in uniaxial cobalt film \cite{exptCo} etc.
 The surface and bulk transition \cite{surcrit} are found to be in
different universality class in the dynamic transition of Ising ferromagnet driven
by oscillating magnetic field. The surface critical
behaviour is observed to differ from that of the bulk in these studies \cite{surcrit}.

 Apart from the Ising model, nonequilibrium dynamic phase transition
 has also been observed in other magnetic models. The off-axial dynamic 
phase transition has been observed in the anisotropic classical Heisenberg model \cite{anisoHei}
 and in the XY model \cite{xy}. The multiple (surface and bulk) dynamic transition 
has been observed \cite{mulHei} in the classical Heisenberg model. The dynamic transition 
has also been observed in the kinetic spin-3/2 Blume-Capel model \cite{BCmodel} and 
in the Blume-Emery-Griffith model \cite{BEGmodel}. To study the dynamical phase 
transition in mixed spin systems also
took much attention in modern research\cite{devi,utem,vatans,ertas,shi}.

Mainly, sinusoidally oscillating or randomly varying magnetic field, 
 which are {\it uniform} over the space (lattice) at any instant of 
time has been used
 to study the nonequilibrium dynamical phase transition and other 
characteristic behaviour 
in Ising magnets and in various other magnetic models. 
The outcome of the above mentioned studies of the nonequilibrium
 phase transition has prompted the researchers towards 
the situations where magnetic excitations are also {\it varied in space} at 
any particular instant of time.
 Propagating magnetic field wave is an example of
 such spatially as well as temporally varied magnetic field applied to the ferromagnetic system.
 This kind of variation of magnetic field is closely related to the situation where an electromagnetic wave passes 
through a magnetic system. Actually, the varying ({\it in time as well as in space}) 
magnetic field wave coupled with the spins of the ferromagnetic system affects
 the dynamic nature of the system. 

The nonequilibrium phase transition in Ising ferromagnet swept by propagating
magnetic field wave is studied\cite{prop}.
 Similar observations are obtained in the random field 
Ising model (RFIM) swept by propagating magnetic field wave \cite{rfimprop}. 
A {\it pinned phase} and a {\it phase of coherent motion of spin clusters} have been observed. 
In RFIM the nonequilibrium phase transition has been studied at zero temperature 
and is tuned by {\it quenched random (field) disorder} \cite{rfimprop}.

Pinned phase and propagating phase (phase of coherent motion of 
spin clusters) are also observed 
in the two dimensional Ising ferromagnet swept by 
propagating magnetic field wave \cite{polwav}. 
The transition is detected by studying the 
{\it variance} of the dynamic order parameter, the {\it derivative} of the 
dynamic order parameter, and the {\it dynamic specific heat} which show sharp 
{\it peak or dip} near transition temperature. In the propagating phase spin clusters 
form a {\it definite pattern} which move coherently with the magnetic field wave, 
whereas in the pinned phase the spin clusters do not move 
coherently in time. The dynamic phase transition is observed to depend upon the 
amplitude and wave length of the propagating magnetic wave. 
The phase boundary is found to {\it shrink} towards the low temperature 
for shorter wavelengths. The relevant length scale also diverges near the transition. 
A dynamic symmetry breaking {\it breathing and spreading} transitions \cite{breath} 
are also recently found in Ising ferromagnet irradiated by spherical magnetic wave.
The nonequilibrium behaviour of the random field Ising ferromagnet, at zero temperature,
 driven by standing magnetic field wave \cite{RFIM} has been studied recently by Monte Carlo simulation in
two dimensions using uniform, bimodal and Gaussain distributions of the
 quenched random fields. Depending on the values of the amplitude of standing magnetic field wave 
and the strength of quenched random field three distinct nonequilibrium phases namely, 
pinned, oscillating spin clusters and random are observed. These phases, though
 have similarities, are different from those found in case of propagating magnetic wave.

There has been much amount of studies done in the nonequilibrium
 dynamic phase transition using the Ising ferromagnet and still
 considerable amount of work is going on to understand other characteristic behaviours 
related to such dynamic phase transitions. But in all these studies 
 in a two dimensional Ising ferromagnet boundary conditions has been kept {\it periodic}, to preserve the translational
invariance. 

It would be interesting to know how the Ising ferromagnet, 
driven by standing magnetic wave behaves at finite temperatures and how the difference
with the propagating magnetic wave can be characterised and quantified. 
How does the boundary affect the dynamic phase transitions? 

In the present study we have shown the effects of Standing magnetic wave on Ising ferromagnet.
 The paper is organised as follows: 
The model and the MC simulation technique are discussed in Sec. II, the numerical 
results are reported in Sec. III and the paper ends with a summary in Sec. IV.

\vskip 0.6cm

\noindent {\bf II. Model and Simulation}
\vskip 0.2cm

The Hamiltonian ({\it time dependent}) of a two dimensional Ising ferromagnet, 
having uniform nearest neighbour spin-spin interaction in presence of 
an external standing magnetic wave is represented by, 
\begin{equation}
 H(t) = -J\Sigma\Sigma' s^z(x,y,t) s^z(x',y',t) - \Sigma h^z(x,y,t)s^z(x,y,t) 
\end{equation}
where $s^z(x,y,t)$ is the Ising {\it spin variable} $(\pm 1)$ at lattice site $(x,y)$ at time $t$. 
The summation  $\Sigma'$ extends over the nearest neighbour sites $(x',y')$ 
of given site $(x,y)$. $J(>0)$ is the {\it ferromagnetic Spin-Spin interaction strength} between the nearest neighbours. 
It is considered to be uniform over the whole lattice for simplicity.
$h^z(x,y,t)$ is the {\it magnetic field} at site $(x,y)$ at time $t$, which has the following form of Standing wave, 
\begin{equation}
 h^z(x,y,t)= h_0 sin (2\pi ft) cos (2\pi x/\lambda)
\end{equation}
The $h_0, f$ and $\lambda$ represent respectively \textit{the field amplitude, the frequency and the wavelength}
 of the standing magnetic wave. Here, the magnetic wave is assumed as {\it linearly polarised} 
along the direction parallel to the spins $(s^z)$. The modulation in amplitude
 of the standing magnetic field wave is considered along the x direction only.
It is worthy to mention that the magnetic field wave considered here is externally applied 
 magnetic field wave and it has no connection with the usual spin wave formed in real ferromagnets.

An $L\times L$ square lattice of Ising spins is considered with 
{\textbf{\textsl{open boundary conditions}}} applied at both directions.
 Also antinodes of standing magnetic field wave are taken at x-boundaries. 
 {\it Monte Carlo Metropolis single spin flip algorithm} 
is used for simulation of the dynamics.
 The initial spin configuration corresponds to high temperature random disordered state 
in which 50\% of the lattice sites have spin state $(+1)$ 
and the other 50\% have $(-1)$.
 Any spin chosen, randomly, at site $(x,y)$ is updated with 
the Metropolis probability \cite{bind} at temperature $T$, given by,   
\begin{equation} 
W(s^z\to -s^z) = Min [exp(-\Delta E/k_BT),1] 
\end{equation}
 where $\Delta E$ is the change in energy due to spin flip  and 
$k_B$ is the Boltzman constant. $L^2$ random updating of spin states in an 
$L\times L$ square lattice constitute the unit time step called 
{\it Monte Carlo Step per Spin} (MCSS). The values of the applied magnetic field and 
the temperature are measured in the units of $J$ and $J/k_B$, respectively. 
Any dynamical state is reached by cooling the system slowly in small steps, 
 from the high-temperature state, which is the dynamically disordered state. 
The values of different dynamical parameters at any temperature are calculated 
after the system achieved steady state and initial transient states are discarded.
The system is kept at constant temperature for a sufficiently long time and the
 average values of those parameters are taken throughout the time
 for consideration of the steady state dynamical behaviour.

\newpage

\noindent {\bf III. Results}
\vskip 0.2cm

In the present study a square lattice of size $(L=100)$ is considered. 
The frequency of standing wave is taken through the study as $f=0.01\ MCSS^{-1}$.
 Different field amplitude $(h_0)$ and wavelength $(\lambda)$ of the standing magnetic field wave
 are considered to study the dependence of transition temperature as dependent on these parameters.
Total length of simulation is $2\times10^5$ MCSS for each temperature value.
 The steady state dynamical behaviour is studied here
 after discarding initial ($5\times10^4$ MCSS) transient data for each temperature value.
 The measured quantities are thus obtained 
 by averaging over $15\times10^4$ MCSS. Since, $f=0.01\ MCSS^{-1}$, a full cycle requires $100\ MCSS$. 
 So, in $15\times10^4\ MCSS$, we have $15\times10^2$ no. of cycles. All dynamical quantities 
are calculated by averaging over $15\times10^2$ cycles.
Temperature is cooled in small steps of $0.05\ J/k_B$, i.e. $\Delta T =0.05\ J/k_B$, here.
 This particular choice is a compromise between the computational time and 
 the precision in measuring the transition temperature.

Two distinct phases namely, {\it Pinned} and {\it Oscillating spin clusters} are identified in the steady state.
The pinned phase is such a phase where all the spins are almost parallel and
remain parallel (along a fixed direction either upward or downward) due to the
small value of the probability of spin flip.
The pinned state is formed below a certain transition temperature called the {\it dynamic transition
temperature}, whereas the oscillating spin clusters phase is formed 
above this temperature. In the low temperature and for small values of the amplitude
of standing magnetic field wave, the probalibility of spin flip becomes very small,
which leads to the dynamical {\it pinned phase}.
These phases are 
shown in fig.1. In the pinned phase most of the 
spins are in some preferred direction i.e. 
either upward or downward but in oscillating spin clusters phase 
approximately half of the total spins are up and the others are
down. The oscillating spin clusters phase has a definite pattern of 
spins forming bands parallel
to $y$ axis. Alternate bands of up and down spins having the bandwidth
$\lambda/2$ are formed in the dynamically disordered phase. The band of up
spins becomes a band of down spins, after a time $1/{2f}$ and it bcemoes again
a band of up spins after a further time interval $1/{2f}$. In this way the
spin band oscillates, forming a standing wave, instead of showing a 
propagation observed in earlier studies\cite{prop}. For sufficiently high values
of temperature and the amplitude of the standing magnetic field wave, due to the
higher rate of spin flip, the system of spins effectively follow the spatio-temporal
variation of applied magnetic standing wave, eventually leading to an {\it oscillating 
spin bands} phase.
 
 The dynamic order parameter $Q$ for such transition is defined as the \textit{time averaged magnetisation 
per site over a full cycle of the standing magnetic field oscillations}, 
i.e. \[Q=\frac{f}{L}\oint\!\!\int\!\! m(x,t)\,dx\,dt.\]
Here $m(x,t)$, defined as, \[m(x,t)=\frac{1}{L}\int\!\!s(x,y,t)\,dy,\] is the
 {\it average instantaneous line magnetisation per site} at lattice coordinate $(x)$, $s(x,y,t)$ being the 
instantaneous spin variable at lattice point $(x,y)$. 
In the pinned phase, the order parameter $Q$ has non-zero value because of arrangement of spins
throughout the whole lattice whereas in the oscillating spin-clusters phase it is zero because spin-clusters are  
arranged in alternate values ($\pm 1$).
So, as temperature decreases $Q$ becomes non-zero (at lower temperature) 
from a zero value (at higher temperature) defining the dynamic transition.

The {\it temperature variations} of the dynamic order parameter $Q$, 
 the derivative of $Q$ i.e. $\frac{dQ}{dT}$, the $L^2$ $\times$ variance of $Q$ i.e. $L^2\langle(\delta Q)^2\rangle$ 
 and the dynamic specific heat $C=\frac{dE}{dT}$,
 where \[E=f\oint \{-J\Sigma\Sigma' s^z(x,y,t) s^z(x',y',t)\}\,dt\] %%% - \Sigma h^z(x,y,t)s^z(x,y,t)
is the average dynamic cooperative energy per spin state of the system (without considering the field energy).
 All the above mentioned dynamical quantities are studied for two different values of field amplitude ($h_0=0.6\ \&\ 1.0$)
 and two different values of wavelength ($\lambda=25\ \&\ 50$) of the standing magnetic field wave 
 (see fig.2 and fig.3).
 The derivatives are calculated numerically using three-point central difference formula \cite{numanaly}.
 All these quantities are calculated statisically over $1500$ different samples (i.e. cycles of standing wave). 
Transition is detected by the sharp peaks (for $L^2\langle(\delta Q)^2\rangle$ and $C$)
 or dip (for $\frac{dQ}{dT}$) in the temperature variations of the corresponding quantities.
 
 It is evident from all the {\it figures} that the transition temperature decreases
 with increase in the field amplitude.  
The nature of transition looks similar to that observed in the case of propagating magnetic field wave \cite{polwav}.
 But there are differences in different phases formed in both the cases. As can be seen in fig.4 
that the instantaneous line magnetisation at lattice sites $(x)$, which lie
  between any two consecutive nodes of standing wave, oscillates coherently with different amplitudes.
 Whereas it propagates along with the propagating wave.
 At nodes of standing magnetic wave, the amplitude of oscillation of instantaneous 
line magnetisation is minimum (zero) and at antinodes it is maximum.
 Thus, the instantaneous line magnetisation at different positions $(x)$ 
forms loops between any two consecutive nodes of standing magnetic field wave.
 In such a case of standing magnetic wave, spins inside a loop oscillate coherently
 where two nearest loops oscillate in opposite phase. At loop boundaries i.e. at nodes spins 
feel minimum effect of the magnetic field and thus their dynamics are more 
thermally driven. On the other hand the dynamics of the spins at the antinodes are governed by the field.
 This is shown in fig.5. The probability that a spin,
 at any site $(x)$ in the lattice will flip, depends on the temperature 
and the local magnetic field strength. Since at nodes the magnetic field strength
is minimum the probability of spin flip is high. The peaks at nodes (fig.5) show that the average
 probability of spin flip over a full period of magnetic oscillation is 
quite large at these sites as compared to other lattice sites. Again in case of Ising ferromagnet,
 driven by propagating magnetic field wave, the probability of spin flip is quite low at all lattice 
sites, since they all feel the same magnetic field strength over a full period.
 This characteristic difference distinguishes between the dynamical phases in Ising ferromagnet 
driven by standing magnetic wave and propagating magnetic wave.

Now collecting all the values of the transition temperatures $T_d$ corresponding to different values of 
the magnetic field amplitudes $h_0$ for a particular wavelength $\lambda$,
 a comprehensive phase boundary may be drawn.
 Fig.6. shows the phase boundary for two different wavelengths $\lambda=25\ \&\ 50.$ As can
be seen from the diagrams that the phase boundary shrinks towards the low field and low 
temperature values for shorter wavelength, which is consistent with the results obtained
previously with propagating and standing magnetic wave using periodic boundary conditions.

\vskip 0.6cm

\noindent{\bf IV. Summary:}
\vskip 0.2cm
 The dynamical response of a two dimensional Ising ferromagnet, having {\it open boundaries}, to the standing 
magnetic field wave is modelled and studied here. MonteCarlo technique is used for simulating 
the observed result. In steady state, two distinct phases; namely {\it pinned} and {\it oscillating spin clusters}
are observed. The pinned phase, with asymmetric and static arrangement of spins 
is the dynamically ordered phase having non-zero value of 
average magnetisation. 
 The oscillating spin clusters phase consists of many parallel band shaped spin clusters and has zero average magnetisation. 
As the system is cooled from high temperature to low temperature, the dynamic order 
parameter becomes non-zero below a certain transition temperature. The dynamic
 transition seems to be of continuous nature and the dynamic transition temperature ($T_d$) depends on the values of the
 amplitude ($h_0$) and the wavelength ($\lambda$) of the standing 
magnetic field wave at a single frequency. It should be mentioned here that the
earlier studies\cite{rev99}, on the dynamic transition, 
in the Ising ferromagnet driven
by oscillating (in time but uniform over space) magnetic field, 
reported the presence of discontinuos transition and located
 a tricritical point
on the phase diagram. Later on, the studies\cite{rev02} on the 
distribution of dynamic order
parameter with much improved statistics showed the absence of any discontinuous
transition. So, to identify any tricritical behaviour (or discontinuous
transition, if any), one should study this with much improved statistics, which
is beyond the scope of our computational facilities. Here, we do not make any
such comment on the presence/absence of any tricritical point. 

 Phase boundaries are drawn for two different wavelengths in $T_d$ versus $h_0$ plane. 
The phase boundary is observed to shrink towards asymetric phase for shorter wavelength. 
 Unlike the propagating phase where instantaneous line magnetisation oscillates with the same amplitude
 at all lattice sites, the same oscillates with different amplitudes at different lattice sites 
along the standing wave. The spins flip more frequently at the nodes of the
 standing wave.
With open boundary condition applied to the lattice the system achieved steady state after longer 
time as compared to periodic boundaries applied to the lattice \cite{polwav}.
 Apart from this, the nature of transition is found similar to the
 earlier studies with periodic boundaries applied in the case of propagating wave.

It would be interesting to see the effects of Standing wave on the highly anisotropic ferromagnetic thin film 
(Co/Ni system) experimentally by time resolved magneto-optic Kerr (TRMOKE) effect.

Controlling the dynamics of a group of spins by external 
magnetic field having a spatio-temporal variation is quite important in the
branch of spintronics, magnonics in modern condensed matter physics\cite{bader}. 
This
present study is a simple statistical mechanical approach of achieving various
dynamical modes of Ising ferromagnet irradiated by a standing magnetic wave, 
just to have a preliminary notion about
the behaviour of a ferromagnetic sample placed in intense optical pattern.

\newpage

\noindent{\bf V. References:}

\vskip 0.6cm
\footnotesize{
}
\newpage

% GNUPLOT: LaTeX picture
\setlength{\unitlength}{0.240900pt}
\ifx\plotpoint\undefined\newsavebox{\plotpoint}\fi
\sbox{\plotpoint}{\rule[-0.200pt]{0.400pt}{0.400pt}}%

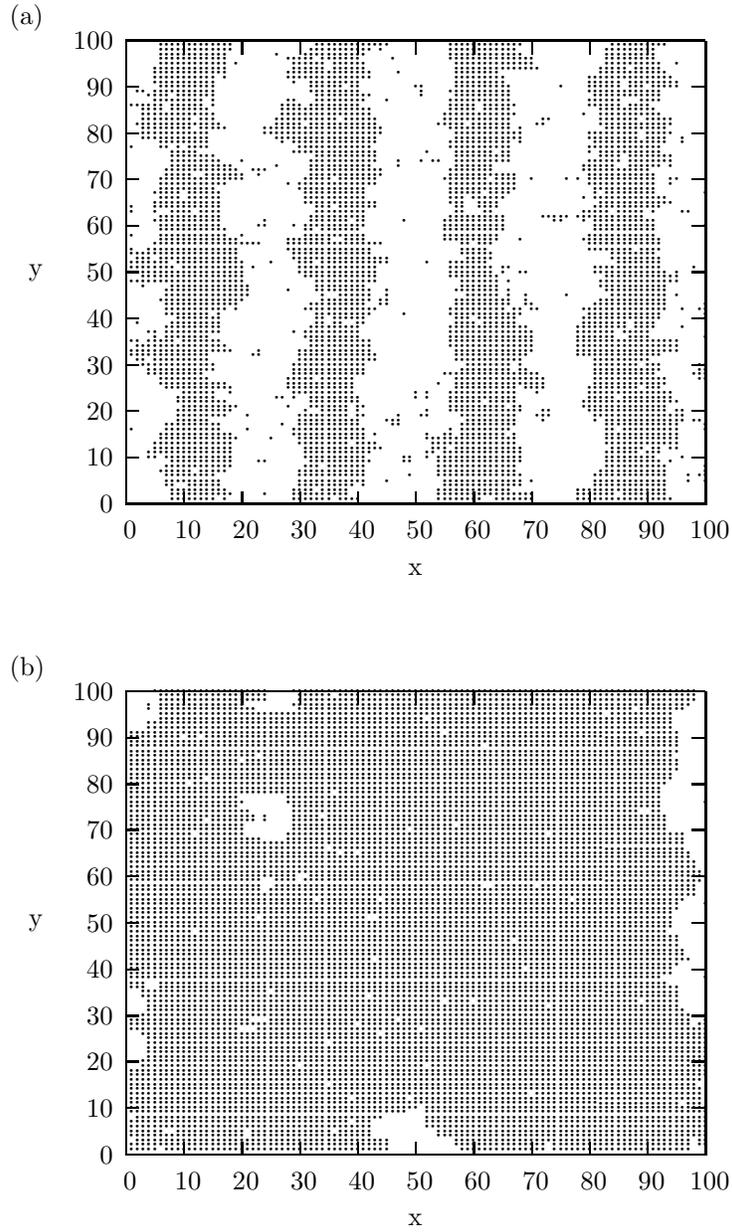
{\\ \footnotesize{{\bf FIG.1.} Dynamical lattice morphology for frequency $f=0.01$  \&\  field amplitude $h_0=0.6$.
			 (a) oscillating spin clusters phase (temp. $T=2.0$), (b) Pinned phase (temp. $T=1.5$). In these figures
 ($\cdot$) symbol denotes the up spin states.}}
\newpage

% GNUPLOT: LaTeX picture
\setlength{\unitlength}{0.240900pt}
\ifx\plotpoint\undefined\newsavebox{\plotpoint}\fi
\sbox{\plotpoint}{\rule[-0.200pt]{0.400pt}{0.400pt}}%
%
\\ {\footnotesize{\bf FIG.2.} Temperature $(T)$ variations of {\bf(a)} $Q$, {\bf(b)} $\frac{dQ}{dT}$, 
{\bf(c)} $L^2\langle(\delta Q)^2\rangle$ and {\bf(d)} $C$ for two different values of standing magnetic 
field amplitude $h_0$. Here $Q$ is the order parameter, $L$ is the lattice size and $C$ is the specific heat.
 Symbols $(\mathbf \bullet)$ \& $(\mathbf\ast)$ represent $h_0=0.6$ \& $h_0=1.0$ respectively.
 The frequency and the wavelength of the standing wave are respectively
$0.01\ MCSS^{-1}$ and $25$ lattice units. The size of the lattice is $100\times 100$.}
\newpage
% GNUPLOT: LaTeX picture
\setlength{\unitlength}{0.240900pt}
\ifx\plotpoint\undefined\newsavebox{\plotpoint}\fi
\sbox{\plotpoint}{\rule[-0.200pt]{0.400pt}{0.400pt}}%

\\ {\footnotesize{\bf FIG.3.} Temperature $(T)$ variations of {\bf(a)} $Q$, {\bf(b)} $\frac{dQ}{dT}$, 
{\bf(c)} $L^2\langle(\delta Q)^2\rangle$ and {\bf(d)} $C_v$ for two different values of standing magnetic 
field amplitude $h_0$. Here $Q$ is the order parameter, $L$ is the lattice size and $C_v$ is the specific heat.
 Symbols $(\mathbf \bullet)$ \& $(\mathbf\ast)$ represent $h_0=0.6$ \& $h_0=1.0$ respectively.
 The frequency and the wavelength of the standing wave are respectively
$0.01\ MCSS^{-1}$ and $50$ lattice units. The size of the lattice is $100\times 100$.}
\newpage
% GNUPLOT: LaTeX picture
\setlength{\unitlength}{0.240900pt}
\ifx\plotpoint\undefined\newsavebox{\plotpoint}\fi
\sbox{\plotpoint}{\rule[-0.200pt]{0.400pt}{0.400pt}}%

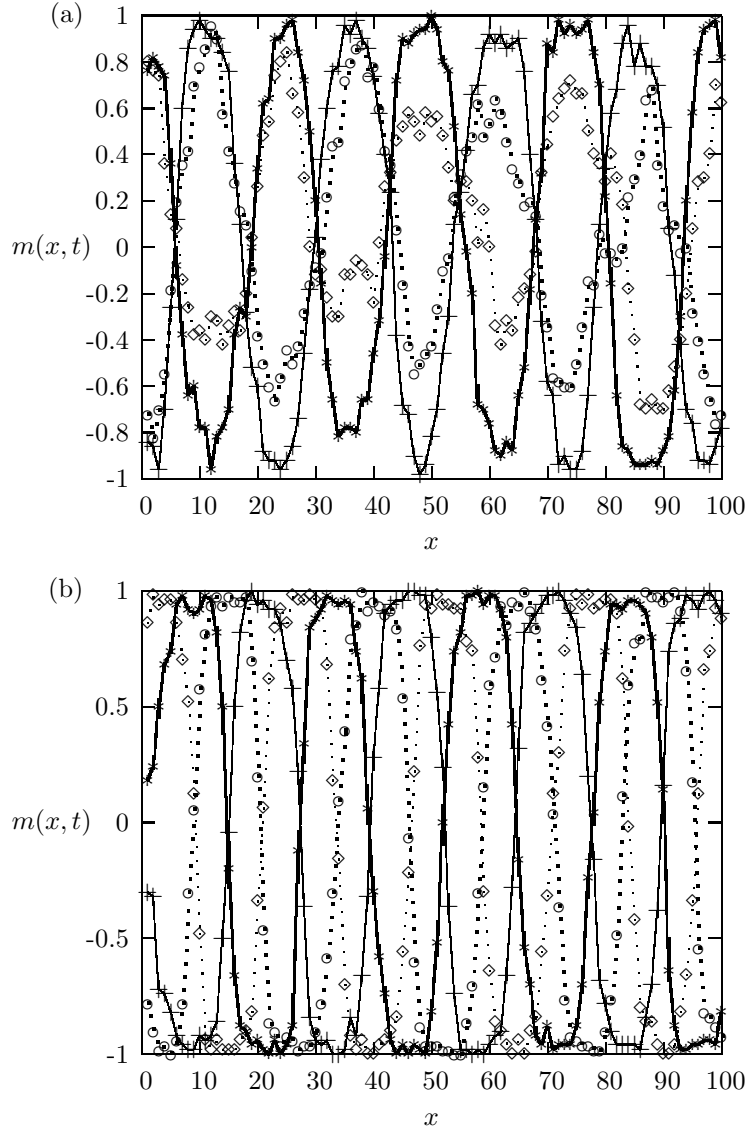
\\ {\footnotesize{\bf FIG.4.} Periodic variation of {\it instantaneous line magnetisation} $m(x,t)$ 
at different lattice sites $(x)$; $(a)$ {\it for standing wave}, $(b)$ {\it for propagating wave}
 in disordered phase, temperature $T=2.00$ in units of $J/k_B$. Different symbols represent
 different times $(+)$ at 199900 MCSS, $(\diamond)$ at 199925 MCSS, $(\ast)$ at 199950 MCSS,
$(\circ)$ at 199975 MCSS, where time period of magnetic field oscillation is 100 MCSS.
 The standing wave is along x axis and the propagating wave propagates along $x$ axis. 
Here the values field amplitude $h_0$, frequency $f$ and wavelength $\lambda$ are $0.6\ J,\ 0.01\ MCSS^{-1}\&
\ 25\ lattice\ units$ respectively.}

\newpage
% GNUPLOT: LaTeX picture
\setlength{\unitlength}{0.240900pt}
\ifx\plotpoint\undefined\newsavebox{\plotpoint}\fi
\sbox{\plotpoint}{\rule[-0.200pt]{0.400pt}{0.400pt}}%

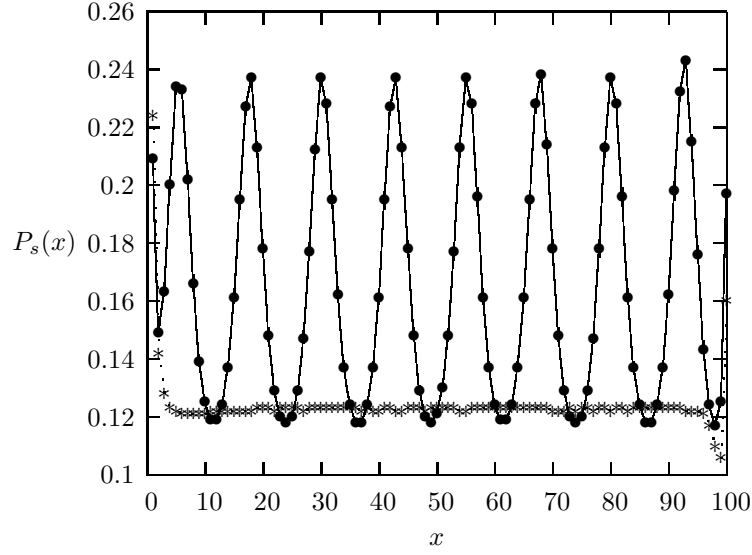
\\ {\footnotesize{\bf FIG.5.} Probability $P_s(x)$ of spin flips at different lattice sites along $x$ axis
 for standing magnetic wave $(\bullet)$ and for propagating magnetic wave $(\ast)$ respectively 
in disordered phase, temperature $T=2.00$ in units of $J/k_B$.
 The standing wave is along $x$ axis and the propagating wave propagates along $x$ axis.
Here the values of field amplitude $h_0$, frequency $f$ and wavelength $\lambda$ are $0.6\ J,\ 0.01\ MCSS^{-1}\&
\ 25\ lattice\ units$ respectively.}
\vskip 0.5 cm
% GNUPLOT: LaTeX picture
\setlength{\unitlength}{0.240900pt}
\ifx\plotpoint\undefined\newsavebox{\plotpoint}\fi
\sbox{\plotpoint}{\rule[-0.200pt]{0.400pt}{0.400pt}}%
\begin{picture}(1500,900)(0,0)
\sbox{\plotpoint}{\rule[-0.200pt]{0.400pt}{0.400pt}}%
\put(350.0,131.0){\rule[-0.200pt]{4.818pt}{0.400pt}}
\put(330,131){\makebox(0,0)[r]{ 0}}
\put(1240.0,131.0){\rule[-0.200pt]{4.818pt}{0.400pt}}
\put(350.0,252.0){\rule[-0.200pt]{4.818pt}{0.400pt}}
\put(330,252){\makebox(0,0)[r]{ 0.5}}
\put(1240.0,252.0){\rule[-0.200pt]{4.818pt}{0.400pt}}
\put(350.0,374.0){\rule[-0.200pt]{4.818pt}{0.400pt}}
\put(330,374){\makebox(0,0)[r]{ 1}}
\put(1240.0,374.0){\rule[-0.200pt]{4.818pt}{0.400pt}}
\put(350.0,495.0){\rule[-0.200pt]{4.818pt}{0.400pt}}
\put(330,495){\makebox(0,0)[r]{ 1.5}}
\put(1240.0,495.0){\rule[-0.200pt]{4.818pt}{0.400pt}}
\put(350.0,616.0){\rule[-0.200pt]{4.818pt}{0.400pt}}
\put(330,616){\makebox(0,0)[r]{ 2}}
\put(1240.0,616.0){\rule[-0.200pt]{4.818pt}{0.400pt}}
\put(350.0,738.0){\rule[-0.200pt]{4.818pt}{0.400pt}}
\put(330,738){\makebox(0,0)[r]{ 2.5}}
\put(1240.0,738.0){\rule[-0.200pt]{4.818pt}{0.400pt}}
\put(350.0,859.0){\rule[-0.200pt]{4.818pt}{0.400pt}}
\put(330,859){\makebox(0,0)[r]{ 3}}
\put(1240.0,859.0){\rule[-0.200pt]{4.818pt}{0.400pt}}
\put(350.0,131.0){\rule[-0.200pt]{0.400pt}{4.818pt}}
\put(350,90){\makebox(0,0){ 0}}
\put(350.0,839.0){\rule[-0.200pt]{0.400pt}{4.818pt}}
\put(532.0,131.0){\rule[-0.200pt]{0.400pt}{4.818pt}}
\put(532,90){\makebox(0,0){ 0.5}}
\put(532.0,839.0){\rule[-0.200pt]{0.400pt}{4.818pt}}
\put(714.0,131.0){\rule[-0.200pt]{0.400pt}{4.818pt}}
\put(714,90){\makebox(0,0){ 1}}
\put(714.0,839.0){\rule[-0.200pt]{0.400pt}{4.818pt}}
\put(896.0,131.0){\rule[-0.200pt]{0.400pt}{4.818pt}}
\put(896,90){\makebox(0,0){ 1.5}}
\put(896.0,839.0){\rule[-0.200pt]{0.400pt}{4.818pt}}
\put(1078.0,131.0){\rule[-0.200pt]{0.400pt}{4.818pt}}
\put(1078,90){\makebox(0,0){ 2}}
\put(1078.0,839.0){\rule[-0.200pt]{0.400pt}{4.818pt}}
\put(1260.0,131.0){\rule[-0.200pt]{0.400pt}{4.818pt}}
\put(1260,90){\makebox(0,0){ 2.5}}
\put(1260.0,839.0){\rule[-0.200pt]{0.400pt}{4.818pt}}
\put(350.0,131.0){\rule[-0.200pt]{0.400pt}{175.375pt}}
\put(350.0,131.0){\rule[-0.200pt]{219.219pt}{0.400pt}}
\put(1260.0,131.0){\rule[-0.200pt]{0.400pt}{175.375pt}}
\put(350.0,859.0){\rule[-0.200pt]{219.219pt}{0.400pt}}
\put(209,495){\makebox(0,0){$h_0$}}
\put(805,29){\makebox(0,0){$T_d$}}
\put(532,374){\makebox(0,0)[l]{($Q\ne0$)}}
\put(896,616){\makebox(0,0)[l]{($Q=0$)}}
\put(1169,131){\usebox{\plotpoint}}
\multiput(1166.72,131.58)(-0.561,0.498){95}{\rule{0.549pt}{0.120pt}}
\multiput(1167.86,130.17)(-53.861,49.000){2}{\rule{0.274pt}{0.400pt}}
\multiput(1111.09,180.58)(-0.751,0.498){93}{\rule{0.700pt}{0.120pt}}
\multiput(1112.55,179.17)(-70.547,48.000){2}{\rule{0.350pt}{0.400pt}}
\multiput(1039.11,228.58)(-0.746,0.498){95}{\rule{0.696pt}{0.120pt}}
\multiput(1040.56,227.17)(-71.556,49.000){2}{\rule{0.348pt}{0.400pt}}
\multiput(966.06,277.58)(-0.762,0.498){93}{\rule{0.708pt}{0.120pt}}
\multiput(967.53,276.17)(-71.530,48.000){2}{\rule{0.354pt}{0.400pt}}
\multiput(893.72,325.58)(-0.561,0.498){95}{\rule{0.549pt}{0.120pt}}
\multiput(894.86,324.17)(-53.861,49.000){2}{\rule{0.274pt}{0.400pt}}
\multiput(838.09,374.58)(-0.751,0.498){93}{\rule{0.700pt}{0.120pt}}
\multiput(839.55,373.17)(-70.547,48.000){2}{\rule{0.350pt}{0.400pt}}
\multiput(766.72,422.58)(-0.561,0.498){95}{\rule{0.549pt}{0.120pt}}
\multiput(767.86,421.17)(-53.861,49.000){2}{\rule{0.274pt}{0.400pt}}
\multiput(712.92,471.00)(-0.498,0.667){69}{\rule{0.120pt}{0.633pt}}
\multiput(713.17,471.00)(-36.000,46.685){2}{\rule{0.400pt}{0.317pt}}
\multiput(675.72,519.58)(-0.561,0.498){95}{\rule{0.549pt}{0.120pt}}
\multiput(676.86,518.17)(-53.861,49.000){2}{\rule{0.274pt}{0.400pt}}
\multiput(621.92,568.00)(-0.498,0.667){69}{\rule{0.120pt}{0.633pt}}
\multiput(622.17,568.00)(-36.000,46.685){2}{\rule{0.400pt}{0.317pt}}
\multiput(584.72,616.58)(-0.561,0.498){95}{\rule{0.549pt}{0.120pt}}
\multiput(585.86,615.17)(-53.861,49.000){2}{\rule{0.274pt}{0.400pt}}
\multiput(530.92,665.00)(-0.495,1.349){33}{\rule{0.119pt}{1.167pt}}
\multiput(531.17,665.00)(-18.000,45.579){2}{\rule{0.400pt}{0.583pt}}
\multiput(512.92,713.00)(-0.495,1.377){33}{\rule{0.119pt}{1.189pt}}
\multiput(513.17,713.00)(-18.000,46.532){2}{\rule{0.400pt}{0.594pt}}
\put(1169,131){\makebox(0,0){$\bullet$}}
\put(1114,180){\makebox(0,0){$\bullet$}}
\put(1042,228){\makebox(0,0){$\bullet$}}
\put(969,277){\makebox(0,0){$\bullet$}}
\put(896,325){\makebox(0,0){$\bullet$}}
\put(841,374){\makebox(0,0){$\bullet$}}
\put(769,422){\makebox(0,0){$\bullet$}}
\put(714,471){\makebox(0,0){$\bullet$}}
\put(678,519){\makebox(0,0){$\bullet$}}
\put(623,568){\makebox(0,0){$\bullet$}}
\put(587,616){\makebox(0,0){$\bullet$}}
\put(532,665){\makebox(0,0){$\bullet$}}
\put(514,713){\makebox(0,0){$\bullet$}}
\put(496,762){\makebox(0,0){$\bullet$}}
\put(605,762){\usebox{\plotpoint}}
\multiput(605,762)(0.000,-20.756){3}{\usebox{\plotpoint}}
\multiput(605,713)(12.453,-16.604){3}{\usebox{\plotpoint}}
\multiput(641,665)(12.507,-16.564){3}{\usebox{\plotpoint}}
\multiput(678,616)(7.288,-19.434){2}{\usebox{\plotpoint}}
\multiput(696,568)(7.157,-19.483){3}{\usebox{\plotpoint}}
\multiput(714,519)(12.453,-16.604){3}{\usebox{\plotpoint}}
\multiput(750,471)(17.233,-11.568){4}{\usebox{\plotpoint}}
\multiput(823,422)(15.638,-13.647){3}{\usebox{\plotpoint}}
\multiput(878,374)(17.233,-11.568){5}{\usebox{\plotpoint}}
\multiput(951,325)(12.453,-16.604){2}{\usebox{\plotpoint}}
\multiput(987,277)(17.233,-11.568){5}{\usebox{\plotpoint}}
\multiput(1060,228)(15.513,-13.789){3}{\usebox{\plotpoint}}
\multiput(1114,180)(15.497,-13.807){4}{\usebox{\plotpoint}}
\put(1169,131){\usebox{\plotpoint}}
\put(605,762){\makebox(0,0){$\ast$}}
\put(605,713){\makebox(0,0){$\ast$}}
\put(641,665){\makebox(0,0){$\ast$}}
\put(678,616){\makebox(0,0){$\ast$}}
\put(696,568){\makebox(0,0){$\ast$}}
\put(714,519){\makebox(0,0){$\ast$}}
\put(750,471){\makebox(0,0){$\ast$}}
\put(823,422){\makebox(0,0){$\ast$}}
\put(878,374){\makebox(0,0){$\ast$}}
\put(951,325){\makebox(0,0){$\ast$}}
\put(987,277){\makebox(0,0){$\ast$}}
\put(1060,228){\makebox(0,0){$\ast$}}
\put(1114,180){\makebox(0,0){$\ast$}}
\put(1169,131){\makebox(0,0){$\ast$}}
\put(350.0,131.0){\rule[-0.200pt]{0.400pt}{175.375pt}}
\put(350.0,131.0){\rule[-0.200pt]{219.219pt}{0.400pt}}
\put(1260.0,131.0){\rule[-0.200pt]{0.400pt}{175.375pt}}
\put(350.0,859.0){\rule[-0.200pt]{219.219pt}{0.400pt}}
\end{picture}
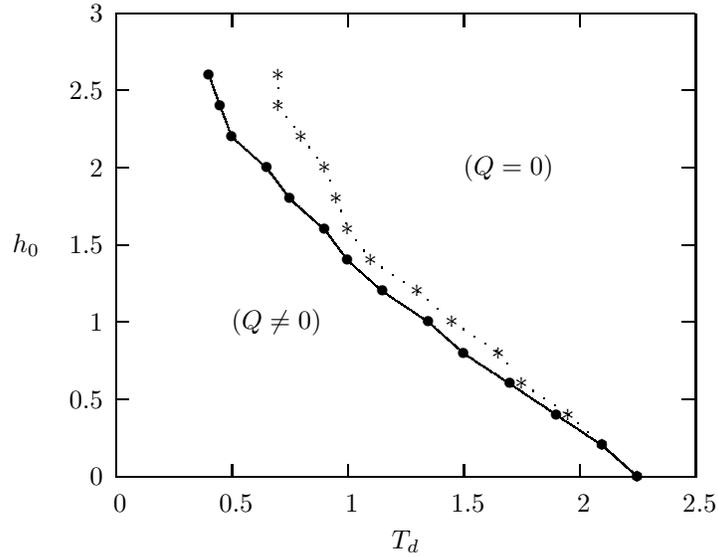
\\ {\footnotesize{\bf FIG.6.} Phase diagram (dynamic transition temperature $T_d$ vs. field amplitude $h_0$)
 for two different wavelength ($\lambda\ = 25\ (\bullet)\ \&\ 50\ (\ast)$) of the standing magnetic field wave.
The frequency of the standing wave is $f=0.01\ MCSS^{-1}$.}

\end{document}